\documentclass[aps,prd,preprint,superscriptaddress,showpacs,byrevtex]{revtex4}
\usepackage{graphicx}
\usepackage{epsfig}
\usepackage{dcolumn}
\usepackage{bm}
\usepackage{amssymb,amsmath}

\newcommand{\etac}{\eta_{c}}
\newcommand{\pp}{\pi^+\pi^-}

\newcommand{\eff}{\varepsilon}
\newcommand{\BR}{{\cal B}}

\newcommand{\pip}{\pi^+}
\newcommand{\pim}{\pi^-}

\newcommand{\etap}{\eta^{\prime}}

\newcommand{\jpsi}{J/\psi}

\newcommand{\EE}{e^+e^-}

\newcommand{\beq}{\begin{equation}}
\newcommand{\eeq}{\end{equation}}
\newcommand{\bitm}{\begin{itemize}}
\newcommand{\eitm}{\end{itemize}}

\begin{document}

\title{\quad\\[0.0cm]\boldmath Search for $CP$ and $P$ violating pseudoscalar decays into $\pi\pi$}

\author{
{\small
M.~Ablikim$^{1}$, M.~N.~Achasov$^{5}$, D.~Alberto$^{38}$, Q.~An$^{36}$, Z.~H.~An$^{1}$, J.~Z.~Bai$^{1}$, R.~Baldini$^{17}$, Y.~Ban$^{23}$, J.~Becker$^{2}$, N.~Berger$^{1}$
% ...
\\
\vspace{0.2cm}
(BESIII Collaboration)\\
\vspace{0.2cm} {\it
$^{1}$ Institute of High Energy Physics, Beijing 100049, P. R. China\\
$^{2}$ Bochum Ruhr-University, 44780 Bochum, Germany\\
% ...
$^{5}$ G.I. Budker Institute of Nuclear Physics SB RAS (BINP), Novosibirsk 630090, Russia\\
% ...
$^{17}$ Laboratori Nazionali di Frascati - INFN, 00044 Frascati, Italy\\
% ...
$^{23}$ Peking University, Beijing 100871, P. R. China\\
% ...
$^{36}$ University of Science and Technology of China, Hefei 230026, P. R. China\\
% ...
$^{38}$ University of Turin and INFN, Turin, Italy\\
% ...
}}
\vspace{0.4cm} }


\date{\today}

\begin{abstract}

  Using a sample of $(225.2\pm 2.8)\times 10^6$ $J/\psi$ events
  collected with the Beijing Spectrometer (BESIII) at the Beijing
  Electron-Positron Collider, $CP$ and $P$ violating decays of $\eta$,
  $\eta'$ and $\etac$ into $\pp$ and $\pi^0 \pi^0$ are searched
  for in $\jpsi$ radiative decays. No significant $\eta$, $\eta'$ or $\etac$ signal is observed,
  and 90\% confidence level upper limits of $\BR(\eta \to
  \pp)<3.9\times 10^{-4}$, $\BR(\etap \to \pp)<5.5\times 10^{-5}$,
  $\BR(\etac \to \pp)<1.3\times 10^{-4}$, $\BR(\eta \to \pi^0
  \pi^0)<6.9\times 10^{-4}$, $\BR(\etap \to \pi^0 \pi^0)<4.5\times
  10^{-4}$ and $\BR(\etac \to \pi^0 \pi^0)<4.2\times 10^{-5}$ are
  obtained.

\end{abstract}


\maketitle

\section{introduction}

Finding the source of $CP$ violation is one of the most important
goals of particle physics. Violation of $CP$ symmetry has important
consequences; it is one of the key ingredients for the
matter-antimatter asymmetry in our universe. $CP$ violation can be
experimentally searched for in processes such as meson decays. The
decays $\eta/\etap/\etac \to \pp$ and $\pi^0 \pi^0$, which violate both
$P$ and $CP$ invariance, provide an excellent laboratory for testing
the validity of symmetries of the physical world. In the Standard
Model (SM), such decays can proceed only via the weak interaction with
a branching fraction of order $10^{-27}$ according to
Ref.~\cite{smpre}.  Higher branching fractions are possible either by
introducing a $CP$ violating term in the QCD lagrangian (a branching
fraction up to $10^{-17}$ can be obtained in this scheme) or allowing
$CP$ violation in the extended Higgs sector (in this case $10^{-15}$
can be reached), as described in Ref.~\cite{smpre}. The detection of
these decays at any level accessible today would signal $P$ and $CP$
violations from new sources, beyond any considered extension of the
SM.

The best previously published results for the $\eta$, $\etap$ and
$\etac$ decays to $\pp$ and $\pi^0 \pi^0$ are: $\BR(\eta \to \pp)<1.3
\times 10^{-5}$~\cite{eta}, $\BR(\etap \to \pp)<2.9\times
10^{-3}$~\cite{etap}, $\BR(\etac \to \pp)<6\times
10^{-4}$~\cite{etac}, $\BR(\eta \to \pi^0\pi^0)<3.5 \times
10^{-4}$~\cite{eta2}, $\BR(\etap \to \pi^0\pi^0)<9 \times
10^{-4}$~\cite{etap2}, and $\BR(\etac \to \pi^0\pi^0)<4 \times
10^{-4}$~\cite{etac} at the 90\% confidence level (C.L.),
respectively.

In this article, results are presented on direct searches for the
decays of $\eta/\etap/\etac \to \pp$ and $\pi^0 \pi^0$ with the BESIII
experiment based on $(225.2\pm 2.8)\times 10^6$ $\jpsi$
events~\cite{jpsinumber}.

\section{BESIII and BEPCII}

BESIII/BEPCII~\cite{bepc2} is a major upgrade of the BESII experiment
at the BEPC accelerator~\cite{bepc1} for studies of hadron
spectroscopy and $\tau$-charm physics~\cite{bes3yellow}. The design
peak luminosity of the double-ring $\EE$ collider, BEPCII, is
$10^{33}$ cm$^{-2}$s$^{-1}$ at a beam current of 0.93 A. The BESIII
detector with a geometrical acceptance of 93\% of 4$\pi$, consists of
the following main components: 1) a small-celled, helium-based main
draft chamber (MDC) with 43 layers. The average single wire resolution
is 135 $\mu$m, and the momentum resolution for 1 GeV/c charged
particles in a 1 T magnetic field is 0.5\%; 2) an electromagnetic
calorimeter (EMC) made of 6240 CsI (Tl) crystals arranged in a
cylindrical shape (barrel) plus two endcaps. For 1.0 GeV photons, the
energy resolution is 2.5\% in the barrel and 5\% in the endcaps, and
the position resolution is 6 mm in the barrel and 9 mm in the endcaps;
3) a Time-Of-Flight system (TOF) for particle identification composed
of a barrel part made of two layers with 88 pieces of 5 cm thick, 2.4
m long plastic scintillators in each layer, and two endcaps with 96
fan-shaped, 5 cm thick, plastic scintillators in each endcap. The time
resolution is 80 ps in the barrel, and 110 ps in the endcaps,
corresponding to a $K/\pi$ separation by more than 2$\sigma$ for momenta
below about 1 GeV/c; 4) a muon chamber system (MUC) made of
1000 m$^2$ of Resistive Plate Chambers (RPC) arranged in 9 layers in
the barrel and 8 layers in the endcaps and incorporated in the return
iron yoke of the superconducting magnet.  The position resolution is about
2 cm.

The estimation of physics backgrounds is performed with Monte Carlo
(MC) simulations. The {\sc GEANT4}-based simulation software {\sc
  BOOST}~\cite{boost} includes the geometric and material description
of the BESIII detectors and the detector response and digitization
models, as well as the tracking of the detector running conditions and
performance. The production of the $\jpsi$ resonance is simulated by
the MC event generator KKMC~\cite{kkmc}, while the decays are
generated by EvtGen~\cite{evtgen} for known decay modes with branching
fractions being set to the PDG~\cite{PDG} world average values, and by
Lundcharm~\cite{lundcharm} for the remaining unknown decays. The
analysis is performed in the framework of the BESIII Offline Software
System ({\sc BOSS})~\cite{boss} which takes care of the detector
calibration, event reconstruction and data storage.

\section {\boldmath Search for $\eta$, $\eta'$ and $\etac$ decays into $\pp$}

To search for $\eta$, $\etap$ and $\etac$ decays into $\pp$ in $\jpsi$
radiative decays, candidate events with the topology $\gamma \pip
\pim$ are selected using the following criteria.
Charged tracks are reconstructed from MDC hits. To optimize the
momentum measurement, we select tracks in the polar angle range
$|\cos\theta|<0.93$ and require that they pass within $\pm$10 cm of
the interaction point in the beam direction and within $\pm$1 cm of
the beamline in the plane perpendicular to the beam. Electromagnetic
showers are reconstructed by clustering EMC crystal energies.
Efficiency and energy resolution are improved by including energy
deposits in nearby TOF counters. Showers identified as photon
candidates must satisfy fiducial and shower-quality requirements.  The
minimum energy is 25 MeV for barrel showers ($|\cos\theta|<0.8$) and
50 MeV for endcap showers ($0.86<|\cos\theta|<0.92$). To exclude
showers from charged particles, a photon must be separated by at least
20$^{\circ}$ from any charged track. EMC cluster timing requirements
suppress electronic noise and energy deposits unrelated to the event.

The TOF (both Endcap and Barrel) and $dE/dx$ measurements for each
charged track are used to calculate $\chi_{PID}^2(i)$ values and the
corresponding confidence levels $Prob_{PID}(i)$ for the hypotheses
that a track is a pion, kaon, or proton, where $i~(i=\pi/K/p)$ is the
particle type. For pion candidates, we require
$Prob_{PID}(\pi)>Prob_{PID}(K)$ and $Prob_{PID}(\pi)>0.001$.

Candidate events must have two charged tracks with zero net charge,
and the number of photons should be greater than or equal to one. The
photon candidate with the maximum energy in the $\EE$ center-of-mass
(C.M.)  frame is taken to be the $\jpsi$ radiative decay photon
($\gamma_{rad}$). At least one charged track must be identified as a
pion. We do a four-constraint (4C) kinematic fit imposing energy and
momentum conservation under the $\jpsi \to \gamma \pi^+ \pi^-$ and
$\jpsi \to \gamma K^+ K^-$ hypotheses and require $\chi^2_{\gamma
  \pp}<30$ and $\chi^2_{\gamma \pp}< \chi^2_{\gamma K^+ K^-}$.

Due to the large branching fraction of $\jpsi \to \rho \pi$,
the decay channel $\jpsi \to \rho \pi \to \pp \pi^0$ constitutes
the main source of  background.  To suppress it
and other possible backgrounds with a $\pi^0$, for candidate events
with two or more photons, all pairings of the radiative photon and the
remaining photons in the event are used to form possible $\pi^0$
candidates, and the pairing with its mass closest to the $\pi^0$
nominal mass is selected.  A clear $\pi^0$ signal is observed.  To
remove it, $m_{\gamma\gamma_{rad}}<0.12$~GeV/$c^2$ or
$m_{\gamma\gamma_{rad}}>0.15$~GeV/$c^2$ is required. The efficiencies
of this requirement are 99\% for $\eta (\etap) \to \pp$ and 96\% for
$\etac \to \pp$.

To suppress the $\jpsi \to \EE$ background and other possible
backgrounds with $\EE$, the requirements of $E_{\pi}^{EMC}<1.2$ GeV
and $|\chi_{dE/dx}(\pi)|<3$ are imposed for both $\pi^+$ and $\pi^-$
candidates. Here $E_{\pi}^{EMC}$ means the deposited energy of pion
candidates in the EMC.  To suppress $\jpsi \to \mu^+ \mu^-$
background, we require $E_{\pi^+}^{EMC}>0.4$ GeV or
$E_{\pi^-}^{EMC}>0.4$ GeV only in $\etac \to \pp$ since its
contamination in the low $\pp$ mass region is very small. This
requirement removes 99.96\% of the $\jpsi \to \mu^+ \mu^-$ background
events, while the efficiency is 62.4\% for $\etac \to \pp$ according
to MC simulations.


To avoid possible bias, we first did the background analyses with the
signal regions ($m_{\pi \pi} =$ 0.53-0.56 GeV/$c^2$ for $\eta \to
\pp$, 0.95-0.97 GeV/$c^2$ for $\etap \to \pp$ and 2.95-3.02 GeV/$c^2$
for $\etac \to \pp$) blinded to check if all the simulated backgrounds
described the data outside of the signal regions.  The same was done
in the analysis of $\eta$, $\eta'$ and $\etac$ decay into $\pi^0
\pi^0$ ($m_{\pi \pi}=$ 0.52-0.57 GeV/$c^2$ for $\eta \to \pi^0 \pi^0$,
0.93-0.98 GeV/$c^2$ for $\etap \to \pi^0 \pi^0$ and 2.95-3.02
GeV/$c^2$ for $\etac \to \pi^0 \pi^0$).  Backgrounds from a number of
potential background channels listed in the PDG~\cite{PDG} are studied
with MC simulations. An inclusive $\jpsi$ MC event sample is also used
to investigate other possible surviving background events.

The main non-peaking backgrounds are from $\jpsi \to \rho \pi$, $\jpsi
\to \mu^+ \mu^-$, $\jpsi \to \EE$, $\jpsi \to a_2(1320) \pi \to \gamma
\pp$, $\jpsi \to b_1(1235) \pi \to \gamma \pp$, $\jpsi \to \pp$,
$\jpsi \to \gamma \sigma /f_2(1270)/f_0(1500)/f_0(1710) \to \gamma
\pp$, and ISR events $\EE \to \gamma_{ISR} \pp$.  The possible peaking
backgrounds are $\jpsi \to \gamma \eta$ with $\eta \to \gamma \pp$ for
$\eta \to \pp$, $\jpsi \to \gamma \etap$ with $\etap \to \gamma \rho^0
\to \gamma \pp$ for $\etap \to \pp$, and $\jpsi \to \gamma \etac$ with
$\etac \to \gamma \pp$ for $\etac \to \pp$.  Branching fractions of
$\jpsi \to \gamma \eta \to \gamma \gamma \pp$ and $\jpsi \to \gamma
\etap \to \gamma \gamma \pp$ have been measured~\cite{PDG}.  After
event selection, the contribution from the decay $\jpsi \to \gamma \eta \to
\gamma \gamma \pp$ to the background is less than 11 events within a
mass region of $\pm 2\sigma$ around the $\eta$ mass peak. Since the
$\pp$ mass distribution is smooth, this background may be
neglected. The contamination from the $\jpsi \to \gamma \etap
\to \gamma \gamma \pp$ background channel will be fixed in the fit below. As
for the possible peaking background $\etac \to \gamma \pp$ for $\etac
\to \pp$, this process is OZI suppressed, unlike $\eta/\eta' \to
\gamma \pi^+ \pi^-$, and can only happen by $c \bar{c}$ annihilation,
and the photon must be from the final state quark.  The branching
fraction for such a process is expected to be very small.  It was
calculated to be $\BR(\etac\to \gamma \pp)=4.5\times 10^{-6}$ in
Ref.~\cite{gao} in the framework of nonrelativistic QCD. After QED
contributions are taken into account, the $\BR(\etac\to \gamma \pp)$
becomes $1.5\times 10^{-6}$. With these calculated branching
fractions, the contribution from this peaking background after event
selection can be neglected.

Figure~\ref{up-data-pippim} shows the $\pp$ invariant mass
distributions of the final candidate events in the $\eta$, $\etap$ and
$\etac$ mass regions after removal of the blinded boxes, where dots
with error bars are data and the dashed histograms are all the
simulated normalized backgrounds. No evident signal is observed, and
the simulated backgrounds describe the data well. For $\eta \to \pp$,
a fit with an $\eta$ signal shape obtained from MC simulation and a
2nd order Chebychev function as the background shape gives $17\pm23$
signal events with a statistical significance of 0.8$\sigma$.  For
$\etap \to \pp$, by fitting the $\pp$ invariant mass spectrum with the
MC determined shape for the $\etap$ signal, the normalized mass
distribution from $\jpsi \to \gamma \etap \to \gamma \gamma \pp$ as
the only peaking background, and a 2nd order Chebychev function for
other backgrounds, $0.1\pm15$ events are obtained with a statistical
significance of 0.1$\sigma$.  For $\etac \to \pp$, a fit with an
acceptance-corrected $\etac$ signal shape obtained from MC simulation
and a 3rd order Chebychev function as the background shape gives
$52\pm35$ signal events with a statistical significance of
1.5$\sigma$.  The fitted results are shown in
Fig.~\ref{up-data-pippim} with solid lines, where the arrows show the
signal mass regions which contain 95\% of the signal according to MC
simulations.

\begin{figure}[htbp]
\includegraphics[height=5.2cm, angle=-90]{fig1a.epsi}\hspace{0.2cm}
\includegraphics[height=5.2cm, angle=-90]{fig1b.epsi}\hspace{0.2cm}
\includegraphics[height=5.4cm, angle=-90]{fig1c.epsi}
\caption{\label{up-data-pippim} (a)-(c): The $\pp$ invariant
mass distributions of the final candidate events in the $\eta$, $\etap$ and $\etac$
mass regions, respectively. The dots with error bars are data,
the solid lines are the fit described in the text, and the
dashed histograms are the sum of all the simulated normalized backgrounds.
The arrows show mass regions
which contain around 95\% of the signal according to MC simulations.
}
\end{figure}

We determine a Bayesian 90\% confidence level upper limit on $N_{sig}$
by finding the value $N^{UP}_{sig}$ such that
$$
\frac{\int_{0}^{N^{UP}_{sig}} \mathcal{L} dN_{sig}} {\int_{0}^{\infty} \mathcal{L} dN_{sig}}=0.90,
$$
where $N_{sig}$ is the number of signal events, and $\mathcal{L}$ is
the value of the likelihood as a function of $N_{sig}$. The upper
limits on the numbers of $\eta$, $\etap$ and $\etac$ are determined to
be 48, 32, and 92, respectively.

\section {\boldmath Search for $\eta$, $\eta'$ and $\etac$ decays into $\pi^0 \pi^0$}

To search for $\eta$, $\etap$ and $\etac$ decays into $\pi^0\pi^0$ in
$\jpsi$ radiative decays, candidate events with the topology $\gamma
\pi^0 \pi^0$ are selected using the following selection criteria. An
event must have 5 or 6 photons and no charged tracks. Here in
selecting photons, the EMC cluster timing requirement is not used.
For the $\eta/\eta' \to \pi^0 \pi^0$ modes, the photon with the
maximum energy is identified as the radiative photon, and all the
other remaining photons in the event are used to form two $\pi^0$
candidates. For $\etac \to \pi^0 \pi^0$, for photons with $E < 0.3$
GeV (a potential radiative photon), all possible two photon pairings
of the remaining photons in the event are used to form two $\pi^0$
candidates. If there is more than one radiative photon candidate
($E<0.3$ GeV), the one that gives the smallest
$|m_{5\gamma}-m_{\jpsi}|$ is used.  The candidate event is chosen as
the photon pairing combination giving the minimum $\chi=
\sqrt{(m_{\gamma \gamma 1}-m_{\pi^0})^2+ (m_{\gamma \gamma
    2}-m_{\pi^0})^2 }$.  Endcap-endcap combinations are of lower
quality and only make up about 0.7\% of all the $\pi^0$ candidates, so
they are removed.  The $\gamma \gamma$ invariant masses are required
to satisfy $|m_{\gamma \gamma}-m_{\pi^0}|< 0.01625~\hbox{GeV}/c^2$
($\sim 2.5 \sigma$) for $\eta(\etap) \to \pi^0 \pi^0$ and $|m_{\gamma
  \gamma}-m_{\pi^0}|< 0.0175~\hbox{GeV}/c^2$ ($\sim 2.5 \sigma$) for
$\etac \to \pi^0 \pi^0$. To improve the $\pi^0 \pi^0$ mass resolution,
especially for the $\eta(\eta') \to \pi^0 \pi^0$ mode, a 4C kinematic
fit under the $\jpsi \to \gamma \pi^0 \pi^0$ hypothesis is done.  To
suppress backgrounds with a $\omega (\to \gamma \pi^0)$, events with
the invariant mass of $\gamma \pi^0$ satisfying $0.72<m_{\gamma
  \pi^0}<0.82$ GeV/$c^2$ are rejected.

The main backgrounds with $\pi^0$ signals are from $\jpsi \to
a_2(1320) \pi \to \gamma \pi^0 \pi^0$, $\jpsi \to b_1(1235) \pi \to
\gamma \pi^0 \pi^0$, $\jpsi \to \omega \pi^0 (\pi^0)$ with $\omega \to
\gamma \pi^0$, and $\jpsi \to \gamma \sigma/
f_2(1270)/f_0(1500)/f_0(1710)/f_4(2050) \to \gamma \pi^0 \pi^0$.
Non-$\pi^0$ background contributions are represented by the normalized
number of events in the sidebands of the two $\pi^0$ mass distributions,
where the two $\pi^0$ mass sidebands are defined as
$0.084<m_{\gamma \gamma}<0.098$~GeV/$c^2$ or $0.168<m_{\gamma
  \gamma}<0.182$~GeV/$c^2$.

\begin{figure}[htbp]
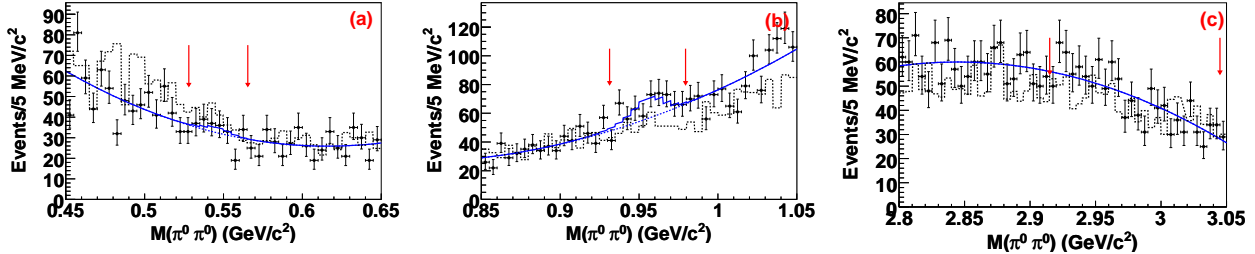

\includegraphics[height=5.2cm, angle=-90]{fig2a.epsi}\hspace{0.2cm}
\includegraphics[height=5.2cm, angle=-90]{fig2b.epsi}\hspace{0.2cm}
\includegraphics[height=5.4cm, angle=-90]{fig2c.epsi}
\caption{\label{up-data-2pi0} (a)-(c): The $\pi^0 \pi^0$  invariant
mass distributions of the final candidate events in the $\eta$, $\etap$
and $\etac$ mass regions, respectively. The dots with error bars are data,
the solid lines are the fit described in the text, and the
dashed histograms are the sum of all the simulated normalized backgrounds.
The arrows show mass regions
which contain around 95\% of the signal according to MC simulations.
}
\end{figure}

Figure~\ref{up-data-2pi0} shows the $\pi^0 \pi^0$ invariant mass
distributions of the final candidate events in the $\eta$, $\etap$ and
$\etac$ mass regions with the blinded boxes removed, where dots with
error bars are data and the dashed histograms are all the simulated
normalized backgrounds. No evident signal is observed and the
simulated backgrounds describe the data well. Since there are no
peaking backgrounds in $\eta/\etap/\etac \to \pi^0 \pi^0$, the
invariant mass distributions are fit with $\eta$, $\etap$ and
acceptance-corrected $\etac$ signal shapes obtained from MC
simulations and 2nd order Chebychev functions as background shapes,
and $11\pm18$, $75\pm30$ and $0.1\pm14$ $\eta$, $\etap$ and $\etac$
signal events, respectively, are obtained with the corresponding
statistical significances of 0.6$\sigma$, $2.6\sigma$ and $0.1\sigma$.
The fitted results are shown in Fig.~\ref{up-data-2pi0} as solid
lines.

In the $\etap \to \pi^0 \pi^0$
  invariant mass distribution, a few events accumulate in the signal
  region. They may from $f_0(980)$ decays since the process $\jpsi \to
  \gamma f_0(980) \to \gamma \pi^0 \pi^0$ is not forbidden, and the
  $f_0(980)$ mass is not very far from the small peak. But since the
  signal significance is only $2.6\sigma$, and in the PWA of $\jpsi
  \to \gamma \pi^0 \pi^0$ $f_0(980)$, no signal is evident or
  needed~\cite{guozj}, the peak may be due to a statistical
  fluctuation. To give a conservative measurement on the upper limit
  and since this was a blind analysis with the background estimated
  before hand, the process $\jpsi \to \gamma f_0(980) \to \gamma \pi^0
  \pi^0$ is not considered in the fit.

\section{Systematic Uncertainties}

The sources of the systematic errors for the branching fraction
measurements are summarized in Table~\ref{totalsys}. The uncertainty is
negligible for pion identification since the identification of only
one of the pions is required. The uncertainty in the tracking
efficiency is 2\% per track~\cite{trkerr}, and is additive. The
uncertainty associated with the kinematic fit is determined to be 2\%
for $\eta/\etap/\etac \to \pp$ using the control sample $\jpsi \to \pp
\pi^0$, while the uncertainty can be neglected for $\eta/\etap/\etac
\to \pi^0 \pi^0$ since we require only a very loose $\chi^2$
requirement (the default value is 200).  The uncertainty due to photon
detection is 1\% per photon. This is determined from studies of photon
detection efficiencies in well understood decays such as $\jpsi \to
\rho^0 \pi^0$ and a study of photon conversion via $\EE \to \gamma
\gamma$. The uncertainty due to $\pi^0$ selection is determined from a
high purity control sample of $\jpsi \to \pp \pi^0$ decays. The
$\pi^0$ selection efficiency is obtained from the change in the
$\pi^0$ yield in the $\pp$ recoiling mass spectrum with or without the
$\pi^0$ selection requirement. The difference of $\pi^0$
reconstruction efficiency between data and MC simulation gives an
uncertainty of 1\% per $\pi^0$~\cite{bes3-gpp}.  The uncertainties
associated with the EMC energy and $|\chi_{dE/dx}|$ requirements are
estimated to be 0.2\% using the control sample $\jpsi \to \pp \pi^0$.
According to the MC simulations, the trigger efficiencies for the
decays $\eta/\etap/\etac \to \pp$ are almost 100\%, and the
uncertainties are neglected. The systematic uncertainties due to the
trigger efficiencies in neutral channels $\eta/\etap/\etac \to \pi^0
\pi^0$ are estimated to be $<$ 0.1\%, based on cross checks using
different trigger conditions.  The background uncertainties are
evaluated by changing the background fitting function and the fitting
range.  Errors on the branching fractions of the $\jpsi \to \gamma
\eta/\eta'/\etac$ are taken from the PDG~\cite{PDG}. The uncertainty
due to the resonance parameters, especially the $\etac$ width, is
estimated by varying the parameters by 1$\sigma$.  Finally the
uncertainty on the number of $\jpsi$ events is
1.3\%~\cite{jpsinumber}. Assuming that all of these systematic error
sources are independent, we add them in quadrature to obtain total
systematic errors as shown in Table~\ref{totalsys}.

\begin{table}[htbp]
\caption{Relative systematic errors (\%) of the decay
branching fractions.} \label{totalsys}
\begin{tabular}{l | c | c}
\hline Source & $\eta/\etap/\etac\to  \pp$ & $\eta/\etap/\etac\to
 \pi^0 \pi^0$
\\\hline
 Tracking & 4.0  & $\cdot \cdot \cdot$\\
 Kinematic fit & 2.0 &  $\cdot \cdot \cdot$\\
 Photon efficiency & 1.0 & 5.0 \\
 $\pi^0$ selection &  $\cdot \cdot \cdot$ & 2.0  \\
 MC statistics & 1.4 & 1.4 \\
 EMC energy and $|\chi_{dE/dx}|$ cuts  & 0.2 &  $\cdot \cdot \cdot$ \\
 Trigger efficiency &  $\cdot \cdot \cdot$ & 0.1 \\
 Background shape & 4.2/6.3/6.6 & 5.6/5.5/8.8 \\
 Branching fractions & 3.1/2.9/24  & 3.1/2.9/24  \\
 Resonance parameter & -/-/8.0 & -/-/9.5 \\
 Number of $\jpsi$ events & 1.3 & 1.3 \\
 \hline
 Sum in quadrature & 7.3/8.6/27 & 8.6/8.5/28\\
 \hline
\end{tabular}
\end{table}

\section{summary}

Since no evident $\eta$, $\etap$ or $\etac$ signal is observed in any
decay mode, we determine upper limits on the branching fractions
$\BR(R \to \pi \pi)$ $(R=\eta, \etap, \etac)$ with the
following formula
$$
\BR(R \to \pi\pi)^{UP}= \frac{N^{UP}_{sig}}{N_{\jpsi} \times \epsilon \times \BR(\jpsi \to \gamma R )},
$$ where $N^{UP}_{sig}$ is the upper limit on the number of observed
events for the signal, $\epsilon$ is the detection efficiency obtained
from MC simulation, where the radiative M1 photon is
distributed according to a $1+\cos^2\theta$ distribution and
multihadronic decays of $\eta/\etap/\etac\to \pi \pi$ are
simulated using phase space distributions,
$N_{\jpsi}$ is the total number of
$\jpsi$ events, $(225.2\pm 2.8)\times 10^6$~\cite{jpsinumber}, which
is obtained from inclusive hadronic decays. For the decay $R \to \pi^0
\pi^0$, this should be divided by the square of the $\pi^0 \to \gamma
\gamma$ branching fraction, when calculating the branching fraction of
$R \to \pi^0 \pi^0$.

Table~\ref{summary2} lists the final results for the upper limits on
the branching fractions of all the processes studied, together with
the upper limits on the numbers of signal events, their detection
efficiencies and signal significances. Here in order to calculate
conservative upper limits on these branching fractions, the
efficiencies have been lowered by a factor of $1-\sigma_{\rm sys}$.
For comparison, we also list the best upper limits on
$\eta/\etap/\etac$ decays to $\pp$ and $\pi^0 \pi^0$ states to date
from the PDG~\cite{PDG}.  Except for $\BR(\eta \to \pi\pi)$,
where the KLOE and GAMS-$4\pi$ Collaborations have huge $\eta$ samples
providing the lowest upper limits for $\eta \to \pp$ and $\eta \to
\pi^0 \pi^0$, our upper limits are the lowest.  These results provide
experimental limits for theoretical models predicting how much $CP$ and
$P$ violation there may be in $\etap$ and $\etac$ meson decays.

\begin{table}[htbp]
\caption{Summary of the limits on $\eta/\etap/\etac$ decays to $\pp$
and $\pi^0 \pi^0$ states. $N^{\rm UP}_{\rm sig}$ is the upper limit on
the number of signal events, $\eff$ is the efficiency, $\sigma_{\rm
sys}$ is the total systematic error, $S$ is the number of statistical
significance, $\BR^{\rm UP}$ is the upper limit at the 90\% C.L. on
the decay branching fraction of $\eta/\etap/\etac$ to $\pp$ or $\pi^0
\pi^0$, and $\BR^{\rm UP}_{PDG}$ is the upper limit on the decay
branching fraction from PDG~\cite{PDG}.} \label{summary2}
\begin{center}
\begin{tabular}{c  c  c  c  c  c  c  }
\hline
 Process  & $N^{\rm UP}_{\rm sig}$~ &
~$\eff$~(\%)~ & $\sigma_{\rm sys} (\%)$   & \hspace{0.2cm} $S$ \hspace{0.2cm}  & $\BR^{\rm UP}$   &$\BR^{\rm UP}_{PDG}$ \\
\hline$\eta \to \pp$ & 48 & 54.28 &  7.3 &  \hspace{0.2cm}0.8$\sigma$ \hspace{0.2cm}  & $3.9\times 10^{-4}$  &$1.3\times 10^{-5}$  \\
$\etap \to \pp$ & 32 & 53.81 & 8.6 &  \hspace{0.2cm}0.1$\sigma$ \hspace{0.2cm} & $5.5\times 10^{-5}$ & $2.9 \times 10^{-3}$ \\
$\etac \to \pp$ & 92& 25.27 &  27 &  \hspace{0.2cm}1.5$\sigma$\hspace{0.2cm} & $1.3 \times 10^{-4}$ &$6 \times 10^{-4}$\\
$\eta \to \pi^0 \pi^0$ & 36 &  23.75 & 8.6 & \hspace{0.2cm} 0.6$\sigma$ \hspace{0.2cm} & $6.9\times 10^{-4}$  &$3.5\times 10^{-4}$  \\
$\etap \to \pi^0 \pi^0$ & 110 & 23.18 &  8.5 & \hspace{0.08cm} 2.6$\sigma$ \hspace{0.2cm}& $4.5\times 10^{-4}$ & $9 \times 10^{-4}$ \\
$\etac \to \pi^0 \pi^0$ & 40 &  35.70 & 28 & \hspace{0.2cm} 0.1$\sigma$ \hspace{0.2cm} &  $4.2\times 10^{-5}$  &$4\times 10^{-4}$\\
 \hline
\end{tabular}
\end{center}
\end{table}

\section{acknowledgments}

The BESIII collaboration thanks the staff of BEPCII and the computing center for their hard efforts. This work is supported in part by the Ministry of Science and Technology of China under Contract No. 2009CB825200; National Natural Science Foundation of China (NSFC) under Contracts Nos. 10625524, 10821063, 10825524, 10835001, 10935007; the Chinese Academy of Sciences (CAS) Large-Scale Scientific Facility Program; CAS under Contracts Nos. KJCX2-YW-N29, KJCX2-YW-N45; 100 Talents Program of CAS; Istituto Nazionale di Fisica Nucleare, Italy; Siberian Branch of Russian Academy of Science, joint project No 32 with CAS; U. S. Department of Energy under Contracts Nos. DE-FG02-04ER41291, DE-FG02-91ER40682, DE-FG02-94ER40823; University of Groningen (RuG) and the Helmholtzzentrum fuer Schwerionenforschung GmbH (GSI), Darmstadt; WCU Program of National Research Foundation of Korea under Contract No. R32-2008-000-10155-0

\end{document}